\documentclass[a4paper]{jpconf}
\usepackage{graphicx}

\usepackage{graphicx}
\usepackage{float}
\usepackage{hyperref}

\usepackage[sorting=none]{biblatex}
\begin{document}
\title{The Virtual Research Environment: towards a comprehensive analysis platform}

\author{Elena Gazzarrini$^1$, Enrique Garcia$^1$, Domenic Gosein$^{1, 2}$, Alba Vendrell Moya$^1$, Agisilaos Kounelis$^{1, 3}$, Xavier Espinal$^1$}
\address{$^1$European Council for Nuclear Research (CERN), 1211 Meyrin, Switzerland}
\address{$^2$Mannheim University of Applied Sciences, Germany}
\address{$^3$Computer Engineering and Informatics Department, University of Patras, Greece}

\ead{elena.gazzarrini@cern.ch}

\begin{abstract}
The Virtual Research Environment is an analysis platform developed at CERN serving the needs of scientific communities involved in European Projects. Its scope is to facilitate the development of end-to-end physics workflows, providing researchers with access to an infrastructure and to the digital content necessary to produce and preserve a scientific result in compliance with FAIR principles. The platform's development is aimed at demonstrating how sciences spanning from High Energy Physics to Astrophysics could benefit from the usage of common technologies, initially born to satisfy CERN’s exabyte-scale data management needs. The Virtual Research Environment’s main components are (1) a federated distributed storage solution (the Data Lake), providing functionalities for data injection and replication through a Data Management framework (Rucio), (2) a computing cluster supplying the processing power to run full analyses with Reana, a re-analysis software, (3) a federated and reliable Authentication and Authorization layer and (4) an enhanced notebook interface with containerised environments to hide the infrastructure’s complexity from the user. The deployment of the Virtual Research Environment is open-source and modular, in order to make it easily reproducible by partner institutions; it is publicly accessible and kept up to date by taking advantage of state of the art IT-infrastructure technologies. 
\end{abstract}

\section{Introduction}
Physicists working at CERN's Large Hadron Collider (LHC) experiments were historically among the first scientists to face large amounts of incoming data, and were therefore forced to find efficient, data-intensive software solutions from an early stage -- the implementation of algorithms for the LHC project started back in the 80's --, much before the `big data' trend emerged on a global scale. Nowadays, PBs of data are saved every day in the CERN Data Center; as an example, the LHCb experiment currently selects 10 GBs of the most interesting LHC collisions each second (after proccessing 4 TBs of data per second in real-time) for physics analysis~\cite{lhc-data}. CERN developers have therefore accumulated experience and expertise in engineering tools for handling, processing and analysing large data volumes.  While High Energy Physics (HEP) sciences have faced these challenges for a long time, non-HEP sciences are more recently entering the exabyte-scale era~\cite{data-astronomy}, and therefore need the ability to efficiently track and process the generated data while meeting FAIR (Findabile, Accessibile, Interoperabile, Reusabile) data principles \footnote{https://www.go-fair.org/fair-principles/}. However, Open Data alone is not sufficient to foster reuse and reproducibility in physics. It is also essential to capture structured information about the analysis workflows to ensure the usability and longevity of results ~\cite{chen2019open, openscience}. A common problem, as stated in the literature~\cite{baker20161}, is that half of the researchers cannot reproduce their own results; this tragic evidence can be alleviated by preserving data and code via (re-)analysis platforms that apply logical techniques to describe, illustrate, condense and evaluate data. EU-funded H2020 projects aim to `democratise' data-intensive technologies, allowing different sciences outside the HEP field -- from High Energy Astrophysics to Gravitational Waves searches -- to gain expertise on new solutions, eventually fostering cross-fertilisation of sciences. All in all, scientific collaborations are becoming more international; as a consequence, common infrastructures that allow reliable and efficient (i) Federated Data Management and Data Transfer Services, (ii) Federated Distributed Storage, (iii) Data Processing and Orchestration and (iv) Software and Analysis Reproducibility are becoming increasingly popular. 
The Virtual Research Environment (VRE) tries to encompass all of the above, while placing special attention on the user experience by providing the scientist with an enhanced notebook interface. The VRE's configuration can in addition be flexibly modified to access heterogeneous external resources (storage and computing) managed by EU partner institutions.  
The following sections will introduce the scientific value of the VRE and illustrate its main components.
\section{Scientific value}

The VRE concept was incubated in the European Science Cluster of Astronomy, Astroparticle and Particle Physics (ESCAPE \footnote{https://projectescape.eu/}) project and is currently being developed and deployed within the EOSC (European Open Science Cloud) Future \footnote{https://eoscfuture.eu/} project, both addressing Open Science challenges to ensure optimised access, management, organisation, processing and preservation of the enormous amount of data handled by the experiments. The tools and concepts initially developed by the ESCAPE work packages, such as the Data Lake (see next section), are hosted and implemented within the VRE, which aims at demonstrating an interdisciplinary science example from bottom-up efforts originating from different scientific domains. In fact, the experiments currently involved in the project come not only from Particle Physics (CERN), but also from High-energy Astrophysics (CTA~\cite{cta}, FermiLAT~\cite{fermilat}), Neutrino Observations (KM3NET~\cite{km3net}, Darkside~\cite{darkside}), Radio Astronomy  (SKA~\cite{skao}, LOFAR~\cite{lofar}) and Gravitational Waves searches (LIGO~\cite{ligo}, Virgo~\cite{virgo}). 

\begin{figure}[h!]
    \centering
    \includegraphics[width=0.75\columnwidth]{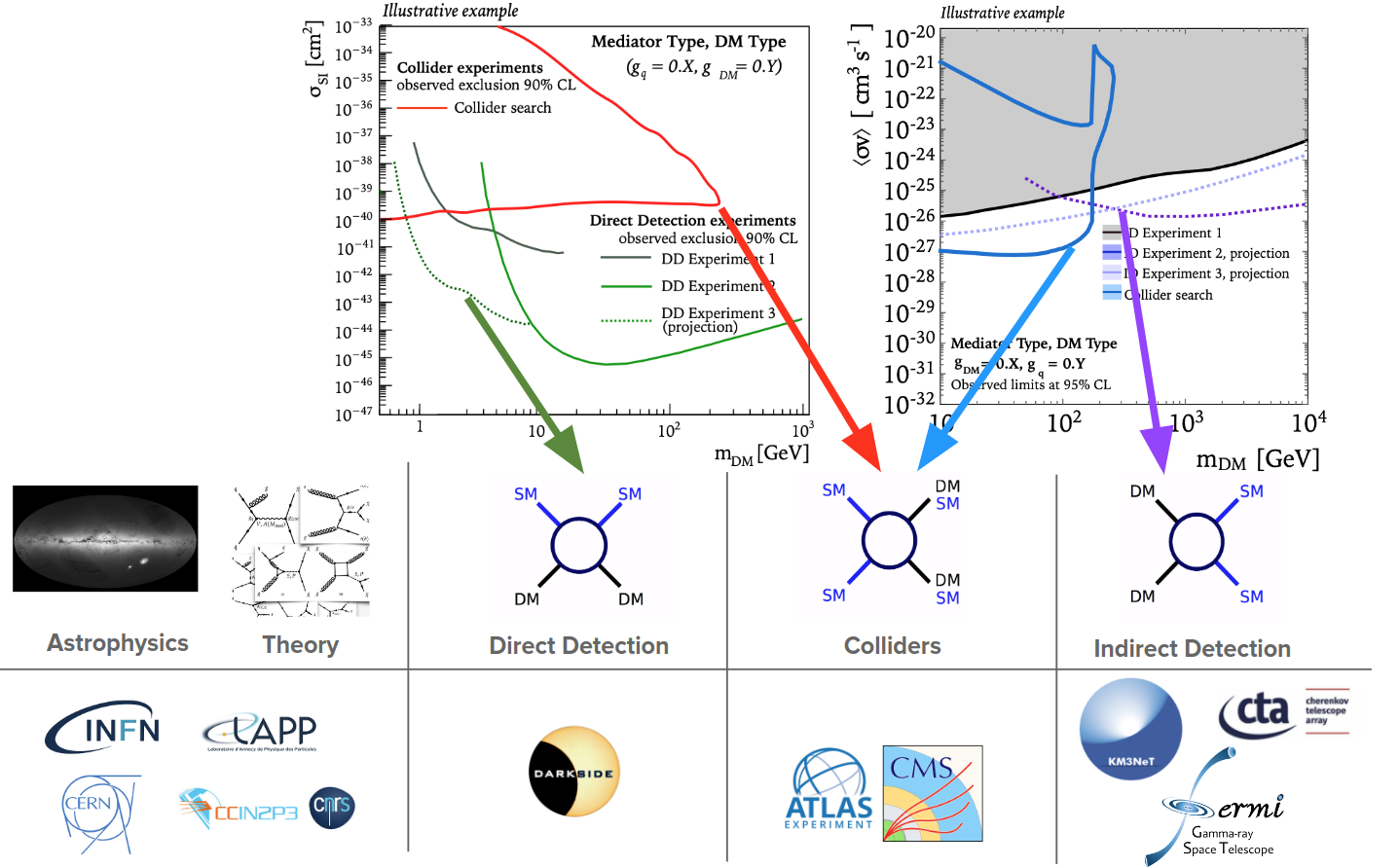}
    \caption{EOSC-Future's Dark Matter Science Project aims at bringing together different search approaches (Astrophysics, Theory, Direct Detection, Collider Physics, Indirect Detection), to ultimately investigate limits on DM mass.}
    \label{fig:DM}
\end{figure}

Many of the aforementioned experiments tackle Dark Matter (DM) exploration from different perspectives. The problem of imposing limits on the mass of DM is a fundamental question in physics: Direct Detection methods study the interaction of particles inside underground detectors, Collider physics produces DM candidates from accelerating protons, Astrophysics observes distant phenomena in the sky and compares them with the theory to detect abnormal behaviours, while Indirect Detection methods investigate annihilating DM by looking at its decay products, such as neutrinos. Figure \ref{fig:DM} illustrates this concept and shows how a platform such as the VRE is a useful place to collect results and to generate combined plots to impose universal DM mass limits. 

\section{VRE Components}

\begin{figure}
    \centering
    \includegraphics[width=0.7\columnwidth]{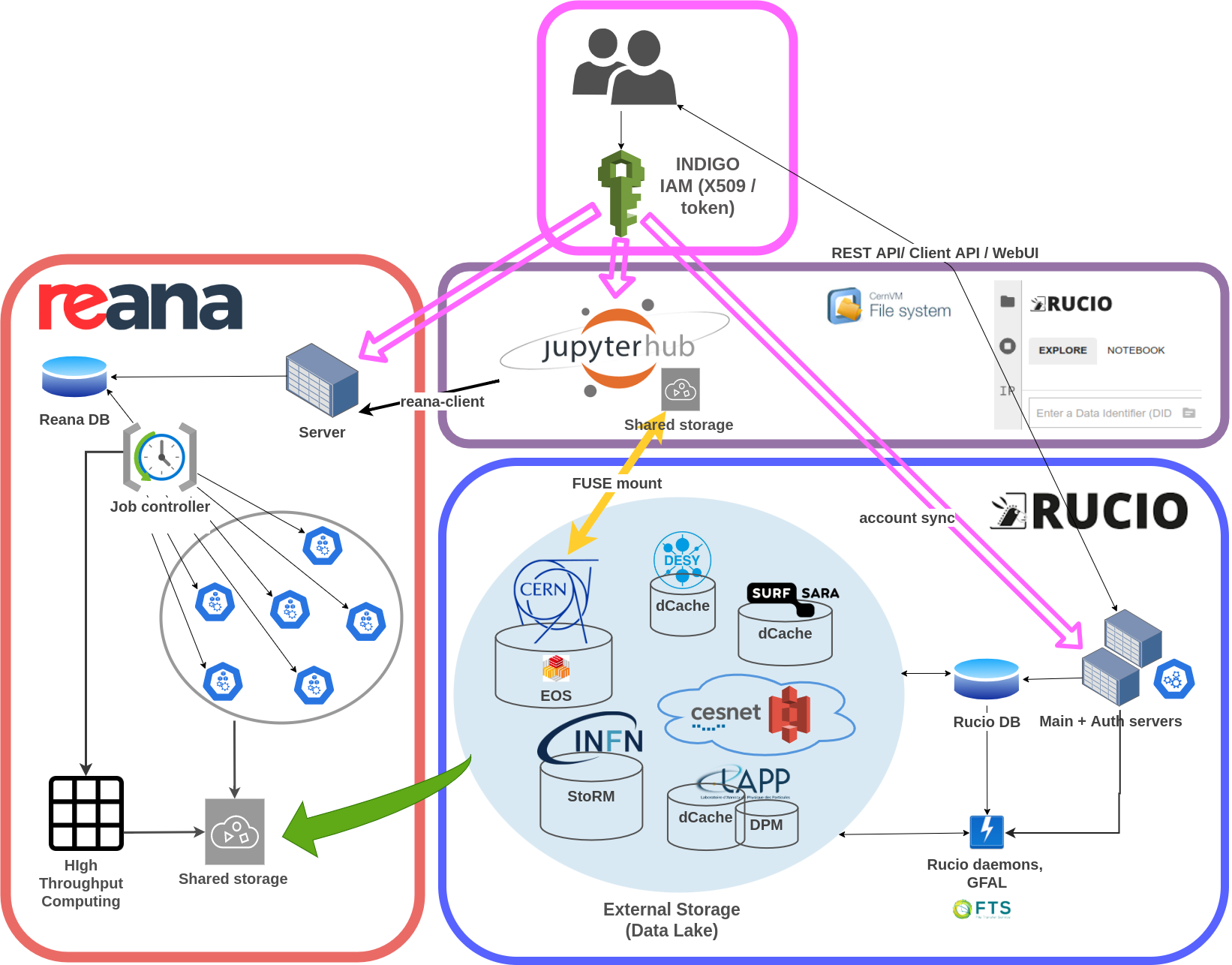}
    \caption{A graphical representation of the VRE components, i.e. (1) a federated distributed storage solution (blue), (2) a computing cluster (red), (3) a federated AAI layer (pink) and (4) an enhanced notebook interface (purple).}
    \label{fig:diagram}
\end{figure}

In its endeavour to homogenise the technological needs of diverse scientific communities, the VRE consists of (1) a data management framework, (2) access to computing processing resources, (3) users management through a reliable Authorisation and Authentication Infrastructure (AAI) and (4) exposure to a user interface to facilitate the interaction with the underlying infrastructure.  Figure \ref{fig:diagram} graphically illustrates the architecture, supported by CI/CD cycles, container orchestration and Infrastructure-as-Code (IaC) processes. The VRE's deployment is centrally managed on CERN's Cloud infrastructure (details can be found in Table \ref{table:tab}) with Kubernetes (K8s). 

\begin{table}[ht]
\centering
\begin{tabular}[t]{|cccccc|}
\hline
vCPUs & RAM (GB) & Masters & Nodes & Remote Storage (TB) & CephFS (TB) \\
\hline
\hline
 184 & 335.8 & 3 & 23 & 646 & 1.8 \\
\hline
\end{tabular}
\caption{The VRE technical components. The first 4 columns refer to the cloud infrastructure managed with K8s, while the last two columns refer to the total quota of the remote storage elements and of the shared object storage (CephFS) attached to the processing nodes. }
\label{table:tab}
\end{table}%

\subsection{Data Management, the Data Lake}

The data management and storage orchestration for the VRE is largely based on a scientific software developed at CERN, Rucio~\cite{rucio2019}. Rucio is an open-source project initially developed by the ATLAS~\cite{ATLAS} experiment for managing community data. It provides services and associated libraries to manage large volumes of data spread across facilities at multiple institutions and organizations. 
The VRE Rucio instance is composed of (i) a cloud infrastructure described in Table \ref{table:tab}, where Rucio servers, daemons and webUI, installed through Helm charts, manage API requests, user authentication, data upload, access, download and replication, (ii) a central relational database hosted at CERN, providing backup services in case of major disruptions and (iii) multiple Rucio Storage Elements (RSEs) with quotas varying from 1 to 300 TB, hosted at each partner institution, supporting various storage technologies (EOS~\cite{eos}, StoRM~\cite{storm}, dCache~\cite{dcache}, DPM~\cite{dpm}, XRootD~\footnote{https://xrootd.slac.stanford.edu/}) and using diverse back-ends (classic RAID systems, Ceph, and multi-replica). Data can be accessed through gridFTP, HTTP(S), XRootD and S3 protocols. Such policy-driven, reliable, distributed data infrastructure, commonly referred to as the Data Lake~\cite{DL1, DL2}, is able to deliver data on-demand at low latency to all types of processing facilities. 


The main functionalities that Rucio offers are (1) data upload, download and streaming, executed by exploiting the power of GFAL~\footnote{https://dmc-docs.web.cern.ch/dmc-docs/gfal2/gfal2.html}, and (2) third-party asynchronous transfers (between RSEs) and deletion, achieved instead by CERN's File Transfer Service (FTS)~\cite{fts}.  The latter is granted permission to access the various RSEs by Rucio processes called daemons, which are responsible for any data management action on the infrastructure. 


\subsection{Computing: Reana cluster} 

The processing of data in the VRE is managed by an instance of CERN's reproducible analysis platform, Reana~\cite{vsimko2019reana}, which allows to run analyses on various computing backends (K8s, HTCondor, Slurm). Navigating the platform is made intuitive for the scientist, who only needs to prepare a declarative \texttt{.yaml} file containing instructions on where to find: (i) input data and parameters, (ii) code, (iii) computing environment and (iv) computational steps needed to perform a full analysis. In this way, scientists can maintain and compare lists of past runs and share the results with colleagues.
Reana's workflow distribution on the cluster's virtual nodes is managed by default by K8s. However, computationally heavier analysis steps can be dispatched to High Performance Computers (HPCs) via HTCondor or Slurm, given the assumption the user has access rights to remote resources. 
The independence of the Reana framework from local storage (when calling input data) was implemented by the VRE team by adding a feature ~\footnote{https://docs.reana.io/advanced-usage/access-control/rucio/} that allows user authentication to the Rucio Data Lake as the first step of the Reana analysis run (represented in Figure \ref{fig:diagram} by a green thick arrow). In this way, data can be moved between the Data Lake's storage elements and the Reana shared storage, located close to the K8s processing nodes where the analysis steps are distributed. 

\subsection{Authentication and Authorization Infrastructure} 
Given the heterogeneous composition of the VRE infrastructure, it is essential to have a single authentication and authorisation method that comprises all services and grants users the correct permissions to access them. The VRE's AAI layer is based on the INDIGO Identity and Access Management (IAM) service~\cite{Ceccanti_2017}. The VRE IAM instance (inherited by the ESCAPE one) is deployed on a K8s cluster at INFN-CNAF and supports authentication via EduGAIN, via OIDC tokens and via X.509 certificates/Virtual Organization Membership Service (VOMS~\footnote{https://github.com/italiangrid/voms}) attribute provisioning services. The token authentication to remote storage elements for data access and transfer -- initially representing the biggest challenge -- has been successfully tested on all the VRE's RSEs. The IAM authentication to the Reana instance is currently being implemented and will appear in the next software release version. 

\subsection{User interface: enhanced notebook service}

The online entry point of the VRE is a JupyterHub interface, where scientists can run preliminary analysis. The user gets authenticated via IAM and selects the desired computational environment, automatically pulled from the VRE container registry. The software is therefore already installed in the Jupyterlab session specific to each user pod. Software distribution services such as the CERN Virtual Machine File System (CVMFS~\footnote{https://cernvm.cern.ch/fs/}) additionally allow software installation on a CephFS 800GB shared volume compatible with POSIX standards, mounted on the Jupyterhub node. 
In order to ensure better data security on the platform and to avoid a user filling up the shared volume (leading to an interruption of the JupyterLab session for all users), the JupyterLab interface has been enhanced with a Rucio plug-in~\footnote{https://github.com/rucio/jupyterlab-extension} (represented in the purple box of Figure \ref{fig:diagram}). This feature enables the user to browse the data in the Data Lake and make a copy of it on a CERN RSE of $\sim$0.5 T, which has been FUSE mounted on the Jupyterhub node (yellow arrow in Figure \ref{fig:diagram}); the data is therefore stored close to the processing power, minimising latency. On the other hand, the Jupyterhub node consists of 14GB of RAM and its usage is limited to an exploratory analysis run; to start larger analyses, it is necessary to connect to the VRE Reana cluster via the terminal of the Jupyterhub and dispatch the computation to distributed HPCs. 

\section{Conclusion}

The modular ecosystem of services and tools constituting the VRE represents an European attempt to demonstrate a bottom-up, FAIR approach to scientific collaboration. The weekly on-boarding of new members requesting an account to access the VRE (with a total of more than 200 users) signifies the community need of such novel infrastructure, that encompasses all the resources needed to easily run an end-to-end analysis. The project has been useful in contributing to improve the software stack of consolidated technologies inside CERN, such as Rucio and Reana. The VRE's applicability to different scientific use cases has been proven successful: postdocs coming from HEP and astrophysics are already using Reana to preserve their workflows on the VRE. The deployment of the infrastructure is kept simple and is extensively documented so it can be used by other institutes as a blueprint: site administrators from collaborations such as the Einstein Telescope and the Deutsches Zentrum für Astrophysik have already demonstrated interest in emulating the VRE at their home institutions. This represents a fundamental achievement under both sociological and technological aspects for European collaborations that should address upcoming data management and computing challenges in the next decade. 

\section*{Code Availability}

The deployment of the VRE infrastructure is still under construction, but the code is available on the public CERN VRE Github project~\footnote{https://github.com/cern-vre}, along with the necessary documentation to reproduce it. The VRE landing page~\footnote{https://escape2020.pages.in2p3.fr/virtual-environment/home/}, provides links to the source codes and description of the various EOSC-Future Science Projects.  

\section*{Acknowledgements}
Authors acknowledge support from the ESCAPE and EOSC-Future projects. ESCAPE has received funding from the European Union’s Horizon 2020 research and innovation programme under Grant Agreement  824064. EOSC-Future is co-funded by the European Union Horizon Programme call INFRAEOSC-03-2020, Grant Agreement 101017536. 

\printbibliography

@online{lhc-data,
  author       = "Cristina Agrigoroae",
  title        = "LHC experiments are stepping up their data processing game",
  url = "https://home.cern/news/news/computing/lhc-experiments-are-stepping-their-data-processing-game",
  urldate = {2022-08-24}
}

@inproceedings{DL1,
  title={ESCAPE Data Lake-Next-generation management of cross-discipline Exabyte-scale scientific data},
  author={Di Maria, Riccardo and Dona, Rizart},
  booktitle={EPJ Web of Conferences},
  volume={251},
  pages={02056},
  year={2021},
  organization={EDP Sciences}
}

@inproceedings{vsimko2019reana,
  title={REANA: A system for reusable research data analyses},
  author={{\v{S}}imko, Tibor and Heinrich, Lukas and Hirvonsalo, Harri and Kousidis, Dinos and Rodr{\'\i}guez, Diego},
  booktitle={EPJ web of conferences},
  volume={214},
  pages={06034},
  year={2019},
  organization={EDP Sciences}
}

@article{baker20161,
  title={1,500 scientists lift the lid on reproducibility},
  author={Baker, Monya},
  journal={Nature},
  volume={533},
  number={7604},
  year={2016}
}

@article{data-astronomy,
  doi = {10.5281/ZENODO.3765389},
  url = {https://zenodo.org/record/3765389}, 
  author = {Rahman, Mubdi and Lang, Dustin and Hlozek, Renee and Bovy, Jo and Perreault-Levasseur, Laurence}, 
  language = {en}, 
  title = {Probing Diverse Phenomena through Data-Intensive Astronomy}, 
  publisher = {Zenodo}, 
  year = {2019},
  copyright = {Creative Commons Attribution 4.0 International}
}

@article{fts,
%doi = {10.1088/1742-6596/513/3/032081},
%url = {https://dx.doi.org/10.1088/1742-6596/513/3/032081},
%year = {2014},
%month = {jun},
%publisher = {},
%volume = {513},
%number = {3},
%pages = {032081},
%author = {A A Ayllon and M Salichos and M K Simon and O Keeble},
%title = {FTS3: New Data Movement Service For WLCG},
%journal = {Journal of Physics: Conference Series},
%abstract = {The File Transfer Service (FTS) is the service responsible for distributing the majority of LHC data across the WLCG infrastructure. We present the current status and features of the new File Transfer Service (FTS3), which addresses the problems that the previous FTS version faced : static channel model, configuration and scalability problems, new protocols support, more database back-ends support, etc. We present the solution we implemented and the design of the new tools as well the reliability, stability, scalability and performance requirements of a data movement middle-ware in the grid environment. The ultimate goal has been to deliver a service which scales horizontally, is easy to install and configure, supports protocols via a plug-in based mechanism (GFAL2) and can perform multi GB/s data transfer on the full mesh of its tiered centres.}
%}

@article{ATLAS,
    author = "Aad, G. and others",
    collaboration = "ATLAS",
    title = "{The ATLAS Experiment at the CERN Large Hadron Collider}",
    doi = "10.1088/1748-0221/3/08/S08003",
    journal = "JINST",
    volume = "3",
    pages = "S08003",
    year = "2008"
}

@online{openscience,
  author       = "Sünje Dallmeier-Tiessen and Tibor Šimko",
  title        = "Open science: A vision for collaborative, reproducible and reusable research",
  url = "https://cerncourier.com/a/open-science-a-vision-for-collaborative-reproducible-and-reusable-research/",
  urldate = {2022-08-31}
}

@article{DL2,
	author = {{Bolton, Rosie} and {Campana, Simone} and {Ceccanti, Andrea} and {Espinal, Xavier} and {Fkiaras, Aristeidis} and {Fuhrmann, Patrick} and {Grange, Yan}},
	title = {ESCAPE prototypes a data infrastructure for open science},
	DOI= "10.1051/epjconf/202024504019",
	url= "https://doi.org/10.1051/epjconf/202024504019",
	journal = {EPJ Web Conf.},
	year = 2020,
	volume = 245,
	pages = "04019",
}

@article{rucio2019,
       author = {{Barisits}, Martin and {Beermann}, Thomas and {Berghaus}, Frank and
         {Bockelman}, Brian and {Bogado}, Joaquin and {Cameron}, David and
         {Christidis}, Dimitrios and {Ciangottini}, Diego and
         {Dimitrov}, Gancho and {Elsing}, Markus and {Garonne}, Vincent and
         {di Girolamo}, Alessandro and {Goossens}, Luc and {Guan}, Wen and
         {Guenther}, Jaroslav and {Javurek}, Tomas and {Kuhn}, Dietmar and
         {Lassnig}, Mario and {Lopez}, Fernando and {Magini}, Nicolo and
         {Molfetas}, Angelos and {Nairz}, Armin and {Ould-Saada}, Farid and
         {Prenner}, Stefan and {Serfon}, Cedric and {Stewart}, Graeme and {Vaand
        ering}, Eric and {Vasileva}, Petya and {Vigne}, Ralph and
         {Wegner}, Tobias},
        title = "Rucio: Scientific Data Management",
      journal = "Computing and Software for Big Science",
       volume = "3",
         year = "2019",
          day = "09",
       number = "1",
        pages = "11",
         issn = "2510-2044",
          doi = "10.1007/s41781-019-0026-3",
          url = "https://doi.org/10.1007/s41781-019-0026-3",
    publisher = "Springer International Publishing"
}

@article{dpm,
author = {Álvarez Ayllón, Alejandro and Beche, Alexandre and Furano, Fabrizio and Hellmich, Martin and Keeble, Oliver and Rocha, Ricardo},
year = {2012},
month = {12},
pages = {2015-},
title = {DPM: Future Proof Storage},
volume = {396},
journal = {Journal of Physics Conference Series},
doi = {10.1088/1742-6596/396/3/032015}
}

@InProceedings{dcache,
author="Fuhrmann, Patrick
and G{\"u}lzow, Volker",
editor="Nagel, Wolfgang E.
and Walter, Wolfgang V.
and Lehner, Wolfgang",
title="dCache, Storage System for the Future",
booktitle="Euro-Par 2006 Parallel Processing",
year="2006",
publisher="Springer Berlin Heidelberg",
address="Berlin, Heidelberg",
pages="1106--1113",
isbn="978-3-540-37784-9"
}

@article{cta,
title = "Design concepts for the Cherenkov Telescope Array CTA: An advanced facility for ground-based high-energy gamma-ray astronomy",
author = "M. Actis and G. Agnetta and F. Aharonian and A. Akhperjanian and J. Aleksi{\'c} and E. Aliu and D. Allan and I. Allekotte and F. Antico and Antonelli, {L. A.} and P. Antoranz and A. Aravantinos and T. Arlen and H. Arnaldi and S. Artmann and K. Asano and H. Asorey and J. B{\"a}hr and A. Bais and C. Baixeras and S. Bajtlik and D. Balis and A. Bamba and C. Barbier and M. Barcel{\'o} and A. Barnacka and J. Barnstedt and {Barres de Almeida}, U. and Barrio, {J. A.} and S. Basso and D. Bastieri and C. Bauer and J. Becerra and Y. Becherini and K. Bechtol and J. Becker and V. Beckmann and W. Bednarek and B. Behera and M. Beilicke and M. Belluso and M. Benallou and W. Benbow and J. Berdugo and K. Berger and T. Bernardino and K. Bernl{\"o}hr and A. Biland and S. Billotta and T. Bird and E. Birsin and E. Bissaldi and S. Blake and O. Blanch and Bobkov, {A. A.} and L. Bogacz and M. Bogdan and C. Boisson and J. Boix and J. Bolmont and G. Bonanno and A. Bonardi and T. Bonev and J. Borkowski and O. Botner and A. Bottani and M. Bourgeat and C. Boutonnet and A. Bouvier and S. Brau-Nogu{\'e} and I. Braun and T. Bretz and Briggs, {M. S.} and P. Brun and L. Brunetti and Buckley, {J. H.} and V. Bugaev and R. B{\"u}hler and T. Bulik and G. Busetto and S. Buson and K. Byrum and M. Cailles and R. Cameron and R. Canestrari and S. Cantu and E. Carmona and A. Carosi and J. Carr and Carton, {P. H.} and M. Casiraghi and H. Castarede and O. Catalano and S. Cavazzani and S. Cazaux and B. Cerruti and M. Cerruti and Chadwick, {P. M.} and J. Chiang and M. Chikawa and M. Cie{\'s}lar and M. Ciesielska and A. Cillis and C. Clerc and P. Colin and J. Colom{\'e} and M. Compin and P. Conconi and V. Connaughton and J. Conrad and Contreras, {J. L.} and P. Coppi and M. Corlier and P. Corona and O. Corpace and D. Corti and J. Cortina and H. Costantini and G. Cotter and B. Courty and S. Couturier and S. Covino and J. Croston and G. Cusumano and Daniel, {M. K.} and F. Dazzi and Angelis, {A. De} and {de Cea Del Pozo}, E. and {de Gouveia Dal Pino}, {E. M.} and {de Jager}, O. and {de La Calle P{\'e}rez}, I. and {de La Vega}, G. and {de Lotto}, B. and {de Naurois}, M. and {de O{\~n}a Wilhelmi}, E. and {de Souza}, V. and B. Decerprit and C. Deil and E. Delagnes and G. Deleglise and C. Delgado and T. Dettlaff and {di Paolo}, A. and {di Pierro}, F. and C. D{\'i}az and J. Dick and H. Dickinson and Digel, {S. W.} and D. Dimitrov and G. Disset and A. Djannati-Ata{\"i} and M. Doert and W. Domainko and D. Dorner and M. Doro and J.-L. Dournaux and D. Dravins and L. Drury and F. Dubois and R. Dubois and G. Dubus and C. Dufour and D. Durand and J. Dyks and M. Dyrda and E. Edy and K. Egberts and C. Eleftheriadis and S. Elles and D. Emmanoulopoulos and R. Enomoto and J.-P. Ernenwein and M. Errando and A. Etchegoyen and Falcone, {A. D.} and K. Farakos and C. Farnier and S. Federici and F. Feinstein and D. Ferenc and E. Fillin-Martino and D. Fink and C. Finley and Finley, {J. P.} and R. Firpo and D. Florin and C. F{\"o}hr and E. Fokitis and Ll. Font and G. Fontaine and A. Fontana and A. F{\"o}rster and L. Fortson and N. Fouque and C. Fransson and Fraser, {G. W.} and L. Fresnillo and C. Fruck and Y. Fujita and Y. Fukazawa and S. Funk and W. G{\"a}bele and S. Gabici and A. Gadola and N. Galante and Y. Gallant and B. Garc{\'i}a and {Garc{\'i}a L{\'o}pez}, {R. J.} and D. Garrido and L. Garrido and D. Gasc{\'o}n and C. Gasq and M. Gaug and J. Gaweda and N. Geffroy and C. Ghag and A. Ghedina and M. Ghigo and E. Gianakaki and S. Giarrusso and G. Giavitto and B. Giebels and E. Giro and P. Giubilato and T. Glanzman and J.-F. Glicenstein and M. Gochna and V. Golev and {G{\'o}mez Berisso}, M. and A. Gonz{\'a}lez and F. Gonz{\'a}lez and F. Gra{\~n}ena and R. Graciani and J. Granot and R. Gredig and A. Green and T. Greenshaw and O. Grimm and J. Grube and M. Grudzi{\'n}ska and J. Grygorczuk and V. Guarino and L. Guglielmi and F. Guilloux and S. Gunji and G. Gyuk and D. Hadasch and D. Haefner and R. Hagiwara and J. Hahn and A. Hallgren and S. Hara and Hardcastle, {M. J.} and T. Hassan and T. Haubold and M. Hauser and M. Hayashida and R. Heller and G. Henri and G. Hermann and A. Herrero and Hinton, {J. A.} and D. Hoffmann and W. Hofmann and P. Hofverberg and D. Horns and D. Hrupec and H. Huan and B. Huber and J.-M. Huet and G. Hughes and K. Hultquist and Humensky, {T. B.} and J.-F. Huppert and A. Ibarra and Illa, {J. M.} and J. Ingjald and Y. Inoue and S. Inoue and K. Ioka and C. Jablonski and A. Jacholkowska and M. Janiak and P. Jean and H. Jensen and T. Jogler and I. Jung and P. Kaaret and S. Kabuki and J. Kakuwa and C. Kalkuhl and R. Kankanyan and M. Kapala and A. Karastergiou and M. Karczewski and S. Karkar and N. Karlsson and J. Kasperek and H. Katagiri and K. Katarzy{\'n}ski and N. Kawanaka and B. Kȩdziora and E. Kendziorra and B. Kh{\'e}lifi and D. Kieda and T. Kifune and T. Kihm and S. Klepser and W. Klu{\'z}niak and J. Knapp and Knappy, {A. R.} and T. Kneiske and J. Kn{\"o}dlseder and F. K{\"o}ck and K. Kodani and K. Kohri and K. Kokkotas and N. Komin and A. Konopelko and K. Kosack and R. Kossakowski and P. Kostka and J. Kotu{\l}a and G. Kowal and J. Kozio{\l} and T. Kr{\"a}henb{\"u}hl and J. Krause and H. Krawczynski and F. Krennrich and A. Kretzschmann and H. Kubo and Kudryavtsev, {V. A.} and J. Kushida and {La Barbera}, N. and {La Parola}, V. and {La Rosa}, G. and A. L{\'o}pez and G. Lamanna and P. Laporte and C. Lavalley and {Le Flour}, T. and {Le Padellec}, A. and J.-P. Lenain and L. Lessio and B. Lieunard and E. Lindfors and A. Liolios and T. Lohse and S. Lombardi and A. Lopatin and E. Lorenz and P. Lubi{\'n}ski and O. Luz and E. Lyard and Maccarone, {M. C.} and T. Maccarone and G. Maier and P. Majumdar and S. Maltezos and P. Ma{\l}kiewicz and C. Ma{\~n}{\'a} and A. Manalaysay and G. Maneva and A. Mangano and P. Manigot and J. Mar{\'i}n and M. Mariotti and S. Markoff and G. Mart{\'i}nez and M. Mart{\'i}nez and A. Mastichiadis and H. Matsumoto and S. Mattiazzo and D. Mazin and McComb, {T. J. L.} and N. McCubbin and I. McHardy and C. Medina and D. Melkumyan and A. Mendes and P. Mertsch and M. Meucci and J. Micha{\l}owski and P. Micolon and T. Mineo and N. Mirabal and F. Mirabel and Miranda, {J. M.} and R. Mirzoyan and T. Mizuno and B. Moal and R. Moderski and E. Molinari and I. Monteiro and A. Moralejo and C. Morello and K. Mori and G. Motta and F. Mottez and E. Moulin and R. Mukherjee and P. Munar and H. Muraishi and K. Murase and Murphy, {A. Stj.} and S. Nagataki and T. Naito and T. Nakamori and K. Nakayama and C. Naumann and D. Naumann and P. Nayman and D. Nedbal and A. Nied{\'z}wiecki and J. Niemiec and A. Nikolaidis and K. Nishijima and Nolan, {S. J.} and N. Nowak and O'Brien, {P. T.} and I. Ochoa and Y. Ohira and M. Ohishi and H. Ohka and A. Okumura and C. Olivetto and Ong, {R. A.} and R. Orito and M. Orr and Osborne, {J. P.} and M. Ostrowski and L. Otero and Otte, {A. N.} and E. Ovcharov and I. Oya and A. Oziȩb{\l}o and S. Paiano and J. Pallota and Panazol, {J. L.} and D. Paneque and M. Panter and R. Paoletti and G. Papyan and Paredes, {J. M.} and G. Pareschi and Parsons, {R. D.} and {Paz Arribas}, M. and G. Pedaletti and A. Pepato and M. Persic and Petrucci, {P. O.} and B. Peyaud and W. Piechocki and S. Pita and G. Pivato and {\L}. P{\l}atos and R. Platzer and L. Pogosyan and M. Pohl and G. Pojma{\'n}ski and Ponz, {J. D.} and W. Potter and E. Prandini and R. Preece and H. Prokoph and G. P{\"u}hlhofer and M. Punch and E. Quel and A. Quirrenbach and P. Rajda and R. Rando and M. Rataj and M. Raue and C. Reimann and O. Reimann and A. Reimer and O. Reimer and M. Renaud and S. Renner and J.-M. Reymond and W. Rhode and M. Rib{\'o} and M. Ribordy and J. Rico and F. Rieger and P. Ringegni and J. Ripken and P. Ristori and S. Rivoire and L. Rob and S. Rodriguez and U. Roeser and P. Romano and Romero, {G. E.} and S. Rosier-Lees and Rovero, {A. C.} and F. Roy and S. Royer and B. Rudak and Rulten, {C. B.} and J. Ruppel and F. Russo and F. Ryde and B. Sacco and A. Saggion and V. Sahakian and K. Saito and T. Saito and N. Sakaki and E. Salazar and A. Salini and F. S{\'a}nchez and {S{\'a}nchez Conde}, {M. {\'A}.} and A. Santangelo and Santos, {E. M.} and A. Sanuy and L. Sapozhnikov and S. Sarkar and V. Scalzotto and V. Scapin and M. Scarcioffolo and T. Schanz and S. Schlenstedt and R. Schlickeiser and T. Schmidt and J. Schmoll and M. Schroedter and C. Schultz and J. Schultze and A. Schulz and U. Schwanke and S. Schwarzburg and T. Schweizer and J. Seiradakis and S. Selmane and K. Seweryn and M. Shayduk and Shellard, {R. C.} and T. Shibata and M. Sikora and J. Silk and A. Sillanp{\"a}{\"a} and J. Sitarek and C. Skole and N. Smith and D. Sobczy{\'n}ska and {Sofo Haro}, M. and H. Sol and F. Spanier and D. Spiga and S. Spyrou and V. Stamatescu and A. Stamerra and Starling, {R. L. C.} and {\L}. Stawarz and R. Steenkamp and C. Stegmann and S. Steiner and N. Stergioulas and R. Sternberger and F. Stinzing and M. Stodulski and U. Straumann and A. Su{\'a}rez and M. Suchenek and R. Sugawara and Sulanke, {K. H.} and S. Sun and Supanitsky, {A. D.} and P. Sutcliffe and M. Szanecki and T. Szepieniec and A. Szostek and A. Szymkowiak and G. Tagliaferri and H. Tajima and H. Takahashi and K. Takahashi and L. Takalo and H. Takami and Talbot, {R. G.} and Tam, {P. H.} and M. Tanaka and T. Tanimori and M. Tavani and J.-P. Tavernet and C. Tchernin and Tejedor, {L. A.} and I. Telezhinsky and P. Temnikov and C. Tenzer and Y. Terada and R. Terrier and M. Teshima and V. Testa and L. Tibaldo and O. Tibolla and M. Tluczykont and {Todero Peixoto}, {C. J.} and F. Tokanai and M. Tokarz and K. Toma and Torres, {D. F.} and G. Tosti and T. Totani and F. Toussenel and P. Vallania and G. Vallejo and {van der Walt}, J. and {van Eldik}, C. and J. Vandenbroucke and H. Vankov and G. Vasileiadis and Vassiliev, {V. V.} and I. Vegas and L. Venter and S. Vercellone and C. Veyssiere and Vialle, {J. P.} and M. Videla and P. Vincent and J. Vink and N. Vlahakis and L. Vlahos and P. Vogler and A. Vollhardt and F. Volpe and {von Gunten}, {H. P.} and S. Vorobiov and S. Wagner and Wagner, {R. M.} and B. Wagner and Wakely, {S. P.} and P. Walter and R. Walter and R. Warwick and P. Wawer and R. Wawrzaszek and N. Webb and P. Wegner and A. Weinstein and Q. Weitzel and R. Welsing and H. Wetteskind and R. White and A. Wierzcholska and Wilkinson, {M. I.} and Williams, {D. A.} and M. Winde and R. Wischnewski and {\L}. Wi{\'s}niewski and A. Wolczko and M. Wood and Q. Xiong and T. Yamamoto and K. Yamaoka and R. Yamazaki and S. Yanagita and B. Yoffo and M. Yonetani and A. Yoshida and T. Yoshida and T. Yoshikoshi and V. Zabalza and A. Zagda{\'n}ski and A. Zajczyk and A. Zdziarski and A. Zech and K. Ziȩtara and P. Zi{\'o}{\l}kowski and V. Zitelli and P. Zychowski",
note = "M1 - Journal Article",
year = "2011",
month = dec,
doi = "10.1007/s10686-011-9247-0",
language = "English",
volume = "32",
pages = "193--316",
journal = "Experimental Astronomy",
issn = "0922-6435",
publisher = "SPRINGER",
number = "3"}

\end{document}